\documentclass[preprint]{aastex}

\def\ngavg{\bar{n}_g}
\def\Navg{N_{\rm avg}}
\def\Mmin{M_{\rm min}}
\def\NNm1{\langle N(N-1) \rangle}
\def\xis{\xi_{\rm 1h}}
\def\xid{\xi_{\rm 2h}}
\def\Rvir{R_{\rm vir}}
\def\intdn{\int_0^\infty dM\frac{dn}{dM}} 
\def\intdnM{\int_0^{M_{\rm max}} dM\frac{dn}{dM}}

\def\hMpc{h^{-1}{\rm Mpc}}

\received{2003 July 1}
\begin{document}

\slugcomment{ApJ, 610, 000 (2004)}
\shortauthors{Zheng}
\shorttitle{Clustering of Red Galaxies at $z\sim 3$}

\title{Interpreting the Observed Clustering of Red Galaxies at $z\sim 3$}
\author{Zheng Zheng}
\affil{Department of Astronomy, Ohio State University, Columbus, OH 43210, 
       USA
      }
\email{zhengz@astronomy.ohio-state.edu}

\begin{abstract}

Daddi et al. have recently reported strong clustering of a population 
of red galaxies at z$\sim$3 in the Hubble Deep Field--South.  
Fitting the observed angular clustering with a power law of index $-0.8$, 
they infer a comoving correlation length $r_0\sim 8 \hMpc$; for a standard 
cosmology, this $r_0$ would imply that the red galaxies reside in rare, 
$M\geq 10^{13} h^{-1}M_\odot$ halos, with each halo hosting $\sim 100$ 
galaxies to match the number density of the population.  Using the 
framework of the halo occupation distribution (HOD) in a $\Lambda$CDM 
universe, we show that the Daddi et al. data can be adequately reproduced 
by less surprising models, e.g., models with galaxies residing in halos of mass 
$M>M_{\rm min}=6.3\times 10^{11} h^{-1}M_\odot$ and a mean occupation 
$\Navg(M)=1.4(M/M_{\rm min})^{0.45}$ above this cutoff. The resultant 
correlation functions do not follow a strict power law, showing instead a clear 
transition from the one-halo--dominated regime, where the two galaxies of 
each pair reside in the same dark matter halo, to the two-halo--dominated 
regime, where the two galaxies of each pair are from different halos. 
The observed high-amplitude data points lie in the one-halo--dominated 
regime, so these HOD models are able to explain the observations despite 
having smaller correlation lengths, $r_0 \sim 5 \hMpc$.  HOD parameters 
are only loosely constrained by the current data because of large sample 
variance and the lack of clustering information on scales that probe the 
two-halo regime. If our explanation of the data is correct, then future 
observations covering a larger area should show that the large scale 
correlations lie below a $\theta^{-1.8}$ extrapolation of the small 
scale points. Our models of the current data suggest that the red galaxies
are somewhat more strongly clustered than UV-selected Lyman-break galaxies
and have a greater tendency to reside in small groups.

\end{abstract}


\section{Introduction}

Clustering of high-redshift galaxies can provide information about
their relation to the underlying mass distribution and their formation 
mechanisms. Efforts have been made to detect high-$z$ galaxies and
to estimate their clustering properties. For example, surveys of 
$z\sim 3$ Lyman break galaxies (LBGs) (\citealt{Adelberger98,Adelberger03,
Steidel98}) show that these galaxies are strongly clustered, with a
correlation length $r_0 \sim 4 \hMpc$. This strong clustering appears
to be naturally explained by theoretical models, which predict high
bias of luminous high-$z$ galaxies (\citealt{Governato98,Kauffmann99,Cen00,
Benson01,Pearce01,Somerville01,Yoshikawa01,Weinberg04}). Recently, using
VLT observations, \citet{Daddi03} have analyzed the clustering properties 
of $K \le 24$ galaxies in the Hubble Deep Field South (HDF-S). They find 
that a population of red galaxies with  $J-K>1.7$ in the photometric 
redshift range $2<z_{\rm phot}<4$ exhibit remarkably strong clustering, 
$r_0 \sim 8 \hMpc$. This paper attempts to interpret these measurements in 
the framework of halo occupation distribution (HOD) models (see, e.g., 
\citealt{Seljak00, Scoccimarro01,Berlind02} and references therein).

Fitting the measured two-point angular correlation function by a power 
law with an index $-0.8$, which corresponds to a power law real-space 
two-point correlation function with an index $-1.8$, \citet{Daddi03} 
derive a correlation length as large as $r_0 \sim 8\hMpc$ for the red 
galaxies. This strong clustering seems hard to reconcile
with conventional models of galaxy bias. For a reasonable cosmological 
model, such as that assumed in the GIF simulation of \citet{Jenkins98}, 
a correlation length of $\sim 8 \hMpc$ corresponds to a linear galaxy 
bias factor of $\sim 5$ at $z\sim 3$. In the halo bias model 
(e.g. \citealt{Mo96}), this bias factor implies that these red galaxies 
would be hosted by $M\ge 10^{13}M_{\odot}$ halos. The comoving number 
density of $M\ge 10^{13}M_{\odot}$ halos is 
$\sim 3\times 10^{-5} h^3{\rm Mpc}^{-3}$. To match the comoving number 
density, $\sim 7\times 10^{-3} h^3{\rm Mpc}^{-3}$, of the red galaxies, 
there should be more than 200 such galaxies in each halo. Even if we take 
into account the fact that galaxy bias is an average of halo bias weighted 
by occupation numbers and lower the halo mass to 
$M\ge 5\times 10^{12} M_{\odot}$, the occupation number is still as
large as about 70. Based on the data, \citet{Daddi03} speculate that 
the problem may be caused by the effect of a small scale excess in the 
correlation function. Detailed modeling is necessary to resolve this 
puzzle.   

For modeling observed galaxy clustering statistics, the framework of the 
HOD is a powerful tool. The HOD describes the relation between the 
distribution of galaxies and that of the matter at the level of individual 
dark matter halos. It characterizes the probability distribution $P(N|M)$ 
that a halo of mass $M$ contains $N$ galaxies of a given type and specifies 
the relative spatial and velocity distributions of galaxies and dark matter 
within halos. With an assumed cosmological model that determines the halo 
population, the HOD can be inferred empirically from observed galaxy 
clustering (\citealt{Peacock00,Marinoni02,Berlind02}). HOD modeling has 
been applied to galaxy clustering data from the Two-Degree Field Galaxy 
Redshift Survey (2dFGRS) and the Sloan Digital Sky Survey (SDSS) (see, e.g., 
\citealt{Bosch03,Magliocchetti03,Zehavi04}). HOD modeling has also been 
used to model the clustering of high-$z$ galaxies, such as LBGs 
(\citealt{Bullock02}) and extremely red objects (EROs) (\citealt{Moustakas02}).

In this paper, we will apply HOD modeling to the population of red galaxies
at $z\sim 3$ in \citet{Daddi03} and try to understand the apparent strong 
clustering of these galaxies. We describe the HOD parameterization and how 
we analytically calculate the galaxy correlation function in \S~2. In \S~3, 
we explain what we learn from model fitting to the observational data. 
We summarize the results and give a brief discussion in \S~4. 

\section{HOD Parameterization and Analytic Calculation of Correlation Function}

Motivated by measurements of the cosmic microwave
background (e.g., \citealt{Netterfield02,Pryke02,Spergel03}),
the abundance of galaxy clusters (e.g., \citealt{Eke96}), and high
redshift supernova observations (e.g., \citealt{Riess98,Perlmutter99}),
we adopt a spatially flat $\Lambda$CDM cosmological model with matter
density parameter $\Omega_m=0.3$ throughout this paper. For the matter
fluctuation power spectrum, we adopt the parameterization of \citet{EBW92}
and assume that the spectral index of the inflationary power spectrum is 
$n_s=1$, the rms fluctuation (linearly evolved to $z=0$) at a scale of 
8$\hMpc$ is $\sigma_8=0.9$, and the shape parameter is $\Gamma=0.21$. 
The Hubble constant is assumed to be $100 h$ km\,s$^{-1}$\,Mpc$^{-1}$ with 
$h=0.7$.   

\subsection{HOD parameterization}

To do an analytical calculation of the correlation function of
galaxies, we need to parameterize the halo occupation distribution.
For the functional form of halo occupation number, we adopt a simple model 
similar to that used by \citet{Zehavi04}, which is loosely motivated by 
results from smoothed particle hydrodynamics (SPH) simulations and 
semi-analytic calculations (see \citealt{Berlind03} and references therein). 
In this model, in halos of mass $M\ge M_1$, the mean occupation number 
follows a power law, $\Navg(M)=(M/M_1)^\alpha$, and in halos of 
mass $M_{\rm min}\le M<M_1$ there is only a single galaxy that is above 
the luminosity threshold, i.e., $\Navg(M)=1$.
Given $\alpha$ and $M_1$, $M_{\rm min}$ is then fully determined by the 
number density of galaxies. Since the correlation function is a statistic
of galaxy pairs, we also need to know the second moment of the
occupation number. SPH simulations and semi-analytic models predict that
the distribution of halo occupation numbers at fixed halo mass is 
much narrower than a Poisson distribution when the occupation is low. 
Here we adopt the so-called nearest-integer distribution for $P(N|\Navg)$, 
which states that the occupation number for a halo of mass $M$ is one of 
the two integers bracketing $\Navg(M)$, with the relative probability 
determined by having the right mean. Besides this basic model, we will also 
consider some alternatives as discussed in \S~3. More detailed discussions
of the parameterization of HOD models appear in \citet{Berlind03} and
\citet{Zehavi04}. Our parameterization here is quite restrictive, but
the data are not sufficient to constrain a model with more freedom.

\subsection{Real Space Correlation Function}

The two-point correlation function of galaxies $\xi(r)$ reflects the 
excess probability over a random distribution of finding galaxy pairs 
with a separation $r$. From the point view of the halo model, the two 
galaxies of each pair can come from either a single halo or two different
halos. Consequently, the two-point correlation function $\xi(r)$ can be
decomposed into two components,
\begin{equation}
\xi(r)=[1+\xis(r)]+\xid(r), 
\end{equation}
where the one-halo term $\xis(r)$ and the two-halo term $\xid(r)$ represent 
contributions by pairs from single halos and different halos, respectively.
The above expression comes from the fact that the total number of galaxy
pairs ($\propto 1+\xi(r)$) is simply the sum of the number of pairs from 
single halos ($\propto 1+\xis(r)$) and that from different halos ($\propto
1+\xid(r)$). The one-halo term and two-halo term dominate respectively at 
small and large separations.

The one-halo term $\xis(r)$ can be exactly computed in real space through
\citep{Berlind02}
\begin{equation}
1+\xis(r)=\frac{1}{2\pi r^2\ngavg^2}
              \intdn\frac{\langle N(N-1)\rangle_M}{2}
              \frac{1}{2\Rvir(M)} F^\prime\left(\frac{r}{2\Rvir}\right),
\end{equation}
where $\ngavg$ is the mean number density of galaxies, $dn/dM$ is the halo 
mass function (\citealt{Sheth99,Jenkins01}), $\langle N(N-1)\rangle_M/2$ is 
the average number of pairs in a halo of mass $M$, and $F(r/2\Rvir)$ is the 
cumulative radial distribution of galaxy pairs, i.e. the average fraction of 
galaxy pairs in a halo of mass $M$ (virial radius $\Rvir$) that have 
separation less than $r$. The function $F^\prime(x)$ depends on the 
profile of the galaxy distribution $\rho_g(r)$ within the halo. In this 
paper, we assume that there is always a galaxy located at the center of the 
halo, and others are regarded as satellite galaxies. With this assumption 
of central galaxies, $F^\prime(x)$ is then the pair-number weighted average 
of the central-satellite pair distribution $F^\prime_{\rm cs}(x)$ and the 
satellite-satellite pair distribution $F^\prime_{\rm ss}(x)$ (see, e.g., 
\citealt{Berlind02,Yang02}),
\begin{equation}
        \frac{\langle N(N-1)\rangle_M}{2}F^\prime(x) =
        \langle N-1\rangle_MF^\prime_{\rm cs}(x) 
      +  \frac{\langle (N-1)(N-2)\rangle_M}{2}F^\prime_{\rm ss}(x).
\end{equation}
The central-satellite galaxy pair distribution $F^\prime_{\rm cs}(x)$ 
is just the normalized radial distribution of galaxies (i.e., 
$\propto \rho_g(r)r^2$), and the  satellite-satellite galaxy pair 
distribution $F^\prime_{\rm ss}(x)$ can be derived through the 
convolution of the galaxy distribution profile with itself 
(see \citealt{Sheth01a}). We will first assume that the galaxy 
distribution is the same as the dark matter distribution within 
the halo, which follows an NFW profile (\citealt{NFW95,NFW96,NFW97}) 
truncated at the virial radius. The concentration of an NFW profile 
depends on the halo mass, for which we use the relation given by 
\citet{Bullock01} after modifying it to be consistent with our halo 
definition -- a gravitationally bound structure with overdensity $\sim 200$. 
Later in this paper, we will also consider a more concentrated galaxy 
distribution profile. 

The two-halo term is basically the average halo correlation function weighted 
by the average occupation number of galaxies of each halo. The halo 
correlation function is related to the mass correlation function by the halo 
bias factor (\citealt{Mo96,Jing98,Sheth01}). It is convenient to calculate
the two-halo term in Fourier space (\citealt{Seljak00,Scoccimarro01}). The
two-halo term contribution to the galaxy power spectrum reads
\begin{equation}
P_{\rm gg}^{\rm 2h}(k)=P_m(k)\left[\frac{1}{\ngavg}\intdnM 
                       \Navg(M) b_h(M) y_g(k,M)\right]^2, 
\end{equation}
where $P_m(k)$ is the mass power spectrum, $\Navg(M)$ is the mean occupation
number in halos of mass $M$, $b_h(M)$ is the halo bias 
factor, $y_g(k,M)$ is the (normalized) Fourier transform of the galaxy 
distribution profile within a halo of mass $M$, and $M_{\rm max}$ is the
upper limit for the integral (see below). In the calculation, we adopt 
the three improvements mentioned in \citet{Zehavi04}. First, for $P_m(k)$,
instead of the linear spectrum as used in previous studies, we use the 
non-linear power spectrum as given by \citet{Smith03} to account for the 
non-linear evolution of the mass (also see \citealt{Magliocchetti03}). 
Second, the halo exclusion effect is approximately considered by choosing 
an appropriate $M_{\rm max}$: for the two-halo term at separation $r$, 
$M_{\rm max}$ is taken to be the mass of the halo with virial radius $r/2$. 
Third, the scale-dependence of the halo bias factor on non-linear scales is 
incorporated by using an empirical formula from simulations. The two-halo term 
of the correlation function is the Fourier transform of the power spectrum,  
\begin{equation}
\xid(r)=\frac{1}{2\pi^2}\int_0^\infty P_{\rm gg}^{\rm 2h}(k) k^2 
        \frac{\sin kr}{kr} dk.
\end{equation}

The correlation function analytically calculated using the above method
agrees fairly well with that measured from a mock galaxy catalog generated 
by populating galaxies according to the same HOD into halos identified 
in $N$-body simulations (see \citealt{Zehavi04} and Figure~2 below).

\subsection{Angular Correlation Function}

The angular distribution of galaxies is a projection of the three-dimensional
distribution. The angular correlation function $w(\theta)$ of galaxies is 
related to the real-space correlation function through Limber's equation
\citep{Peebles80}. In a flat universe, as adopted in this paper, for the 
small-angle limit, Limber's equation has the form
\begin{equation}
w(\theta) = \frac{\int_{r_{\rm min}}^{r_{\rm max}} \ngavg^2(x) S^2(x) x^4 dx
                  \int_{-\Delta r}^{\Delta r} \xi(\sqrt{y^2+x^2\theta^2},z) dy 
                 }{\left[\int_{r_{\rm min}}^{r_{\rm max}} \ngavg(x) S(x) x^2 dx
                   \right]^2
                  },
\end{equation}
where $\Delta r=r_{\rm max}-r_{\rm min}$ is the radial range of the survey,
$\ngavg(r)$ is the average number density of galaxies at distance $r$, and 
$S(r)$ is the selection function of the sample (all distances are in comoving
units). The sample selection function $S(r)$ can be derived from the observed 
redshift distribution if the sample is large enough, but the 49 galaxy sample 
of \citet{Daddi03} is not large enough to allow  a precise measurement.
However, it turns out that the basic result of this paper is not sensitive 
to the form of the selection function. We therefore simply assume that $S(r)$
is constant over the redshift range $z=2$ to $z=4$, which defines 
$r_{\rm min}$ and $r_{\rm max}$.

In practice, the angular correlation function is estimated by comparing 
the observed pair numbers in an angular separation bin with those from 
a random sample of the same geometry. The widely used estimator proposed by
\citet{Landy93} estimates the angular correlation function as
\begin{equation}
w_b(\theta)=\frac{{\rm DD-2DR+RR}}{\rm RR},
\end{equation}
where DD, DR, and RR represent number counts of data-data (galaxy-galaxy) 
pairs, data-random (galaxy-random) pairs, and random-random pairs, 
respectively, in the angular bin around $\theta$. Each of these number counts 
are 
normalized so that the summation over all $\theta$ is unity (i.e., the number
of pairs in each angular bin is divided by the total number of pairs in the
field). The estimated angular correlation function $w_b(\theta)$ is subject 
to a statistical bias that leads to systematically lower values than the
true angular correlation function $w(\theta)$, 
$w_b(\theta)=w(\theta)-I.C.$, where 
\begin{equation}
I.C. = \frac{1}{\Omega^2}\int\int w(\theta) d\Omega_1 d\Omega_2
\end{equation}
is the integral constraint \citep{Groth77}. Since $w_b(\theta)$ is the 
quantity directly measured from the observation, it is more appropriate 
to try to fit $w_b(\theta)$ than to fit $w(\theta)$. To convert the 
analytically predicted $w(\theta)$ to $w_b(\theta)$, we use the 
random-random sample to calculate the integral constraint expected for 
the model $w(\theta)$ (see, e.g., \citealt{Roche99}),
\begin{equation}
I.C. = \frac{\Sigma N_{\rm rr}(\theta) w(\theta)}{\Sigma N_{\rm rr}(\theta)},
\end{equation}
where $N_{\rm rr}(\theta)$ is the count of random-random pairs in
the angular bin around $\theta$. We only need to estimate 
$f(\theta)=N_{\rm rr}(\theta)/\Sigma N_{\rm rr}(\theta)$ once from a 
random sample that has the same geometry as the observation. We generate 
100 such random samples with 5,000 points in each and take the average 
$f(\theta)$ as the estimate.

\section{Fitting the Observations}
\begin{figure}[h] 
\plotone{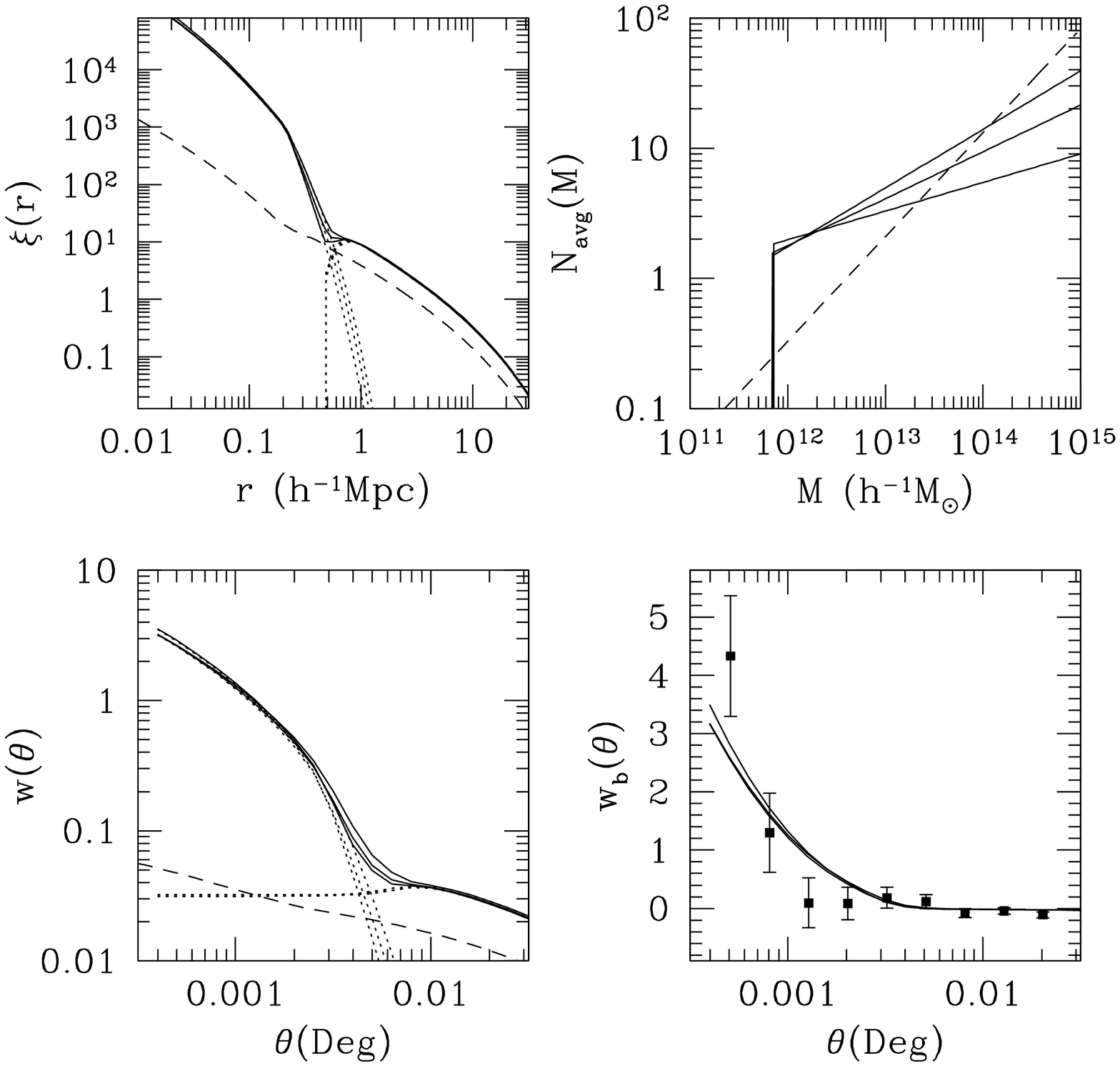}
\caption[]{Illustration of the loose constraints on individual HOD 
parameters for the nearest-integer model. Three cases of parameter 
combinations are shown:
$(M_{\rm min},M_1,\alpha) 
=(6.5\times 10^{11}, 4.5\times 10^{10}, 0.22)$,
$(6.2\times 10^{11}, 2.0\times 10^{11}, 0.36)$, and 
$(6.3\times 10^{11}, 2.9\times 10^{11}, 0.45)$, where masses are in units of
$h^{-1}M_{\odot}$. The top right panel shows the corresponding mean 
occupation number as a function of halo mass for the three cases.  
The top left panel, the bottom left panel, and the bottom right panels are 
for the real space two-point correlation function, the angular correlation 
function, and the measured angular correlation function (i.e., with the 
integral constraint subtracted), respectively. The dotted lines show 
contributions from one-halo pairs and two-halo pairs. Data points with error 
bars in the bottom right panel are from \citet{Daddi03}. The dashed line in
the top right panel illustrates the mean occupation function of the LBGs, 
and the dashed curves in the left panels are corresponding correlation 
functions (see the text).
}
\end{figure}
The angular clustering data we are interested in are for a population
of $K-$selected galaxies ($K\le 24$) at $2<z_{\rm phot}<4$ with $J-K>1.7$. 
Details about this sample can be found in \citet{Daddi03}. The sample includes 
49 galaxies found within a field of view of $\sim$4 arcmin$^2$.
The comoving number density of these galaxies is $\sim 7.1\times 10^{-3} 
h^3{\rm Mpc}^{-3}$. Assuming the angular correlation function to 
be a power law with an index $-0.8$,  \citet{Daddi03} find its amplitude 
at $1^{\circ}$ to be $39.1\pm 10.2$, which corresponds to a real space 
correlation length of $8.3\pm 1.2 \hMpc$ (comoving). 

We now fit the data (kindly provided in electronic form by E. Daddi) 
using the model in \S~2. For a given assumption about
$P(N|\Navg)$, e.g., a nearest-integer or Poisson distribution,
the analytic angular correlation model discussed in \S~2.3 has two free 
parameters: $M_1$, which determines the amplitude of $\Navg(M)$, and $\alpha$, 
which is the slope of $\Navg(M)$ at high halo masses. $M_{\rm min}$ is fixed 
by the mean number density of galaxies in the sample. We thus perform a 
two-parameter $\chi^2$ fit to the data. The observational error bars reported
by \citet{Daddi03} are used in the calculation of $\chi^2$ and are assumed 
to be uncorrelated. With these data and error bars, we find that the two 
free HOD parameters are highly degenerate and that they can be only loosely 
constrained individually. For example, with the nearest-integer distribution, 
$M_1$ is in the range 4--30$\times10^{10} h^{-1}M_{\odot}$ and $\alpha$ is 
in the range 0.2--0.5. If we assume a Poisson distribution for $P(N|\Navg)$, 
then the resultant $\alpha$ is unrealistically large ($\sim 3$), with large 
uncertainty. With the Poisson distribution, if we change the functional form 
of $\Navg(M)$ to be a power law with a low mass cutoff, $\alpha$ can be in 
a reasonable range but still with large uncertainty. Through studying subhalos
in high-resolution dissipationless simulations, \citet{Kravtsov04} propose an 
HOD form, which separates contributions from central and satellite galaxies. 
The mean occupation function of central galaxies is a step function, while
the distribution of satellite galaxies can be approximated by a Poisson 
distribution with the mean following a power law. The resultant shape of
the mean occupation function and scatter around the mean are somewhat similar
to our basic model. We also try this HOD form and find results and
uncertainties similar to those of our basic model.  

We illustrate the looseness of the HOD parameter constraints in Figure~1 by 
showing the results of different parameter combinations that lead to similar 
real-space correlations and angular correlations for the nearest-integer 
case. Note that since $M_{\rm min} > M_1$ is derived from the fit, the 
resultant $\Navg(M)$ is equivalent to a case in which $\Navg(M)$ is a power 
law with a low-mass cutoff, with no ``flat" portion at $M<M_1$. The result 
of $M_{\rm min} > M_1$ mimics the case of local giant elliptical galaxies and 
$z\sim 1$ EROs as modeled by \citet{Moustakas02}, a point discussed further 
at the end of this section. There is a strong break in the modeled 
correlation function between the one-halo and two-halo regimes. The sharpness 
of this break is somewhat exaggerated by the approximate nature of our 
correction for halo exclusion. However, the angular correlation function 
is less affected by this approximation because it is a projection of the 
real-space correlation, and we show later that the approximate treatment 
of halo exclusion has a negligible effect on our $w_b(\theta)$ modeling here.

The reduced $\chi^2$ (7 degrees of freedom, 9 data points, and 2 free 
parameters) from either the nearest-integer model or the power law 
Poisson model is about 1.8, which does not seem to be a good model fit.
The field of view of the survey is less than 4 arcmin$^2$, and the total
number of galaxies is only about 50. Thus, as noted by \citet{Daddi03},
the sample variance may be large, and error bars based only on the finite 
number of objects (as used above) may be too small. We therefore attempt
to make more realistic error estimates by populating halos from the GIF 
simulation (\citealt{Jenkins98}) with galaxies. We use the halo population 
from the GIF simulation output at $z=2.97$ and proceed as follows. 
First, each halo is assumed to have a truncated NFW density profile with 
the same concentration-mass relation used in the analytic model. We then 
populate galaxies according to the halo occupation distribution from the 
$\chi^2$ fitting and generate a mock galaxy catalog. Next, we randomly 
extract 10 slices along one direction from the mock catalog, with the 
cross-section of each slice having the same size and geometry as the 
observation. These 10 slices are checked at selection to make sure that 
they do not overlap (even partially) with each other. The 10 slices are 
stacked together to approximate the radial extent of the survey in comoving 
distance. The projection of the stacked slice thus represents one 
``observation."  Finally, we estimate $w_b(\theta)$ for this observation
in the same angular bins as the real data using the technique of 
\citet{Landy93} (Equation 7). The data-random and random-random terms are 
averaged over 100 random realizations, and each random sample realization 
has 5000 points. Altogether we make 100 observations and estimate 
$w_b(\theta)$ for each one. 

The result is shown in the top left panel of Figure~2. The central solid 
line is the average over the 100 observations, which agrees with the model 
prediction ({\it dot-dashed line}) as expected (and verifying that our 
analytic approximation is accurate enough for our purposes). The dashed 
lines above and below the solid line represent the $1\sigma$ scatter of the 
100 observations. The estimated angular correlation $w_b(\theta)$ for 
an individual observation is very uncertain and may even not decrease
monotonically with $\theta$ (as is the case for the real data points). 
Compared with the scatter derived here, the observational error bars are 
apparently underestimated by a factor of about 1.5. If we take the mock 
catalog scatter as true error bars, then our model fit is acceptable. 
However, the uncertainties in HOD parameters were large even with the 
original error bars, so we do not wish to place much weight on the particular 
values that emerge in the best fit. Rather, we wish to use our HOD models as 
a general guide in understanding the implications of the data.
\begin{figure}[h]
\epsscale{.90}
\plotone{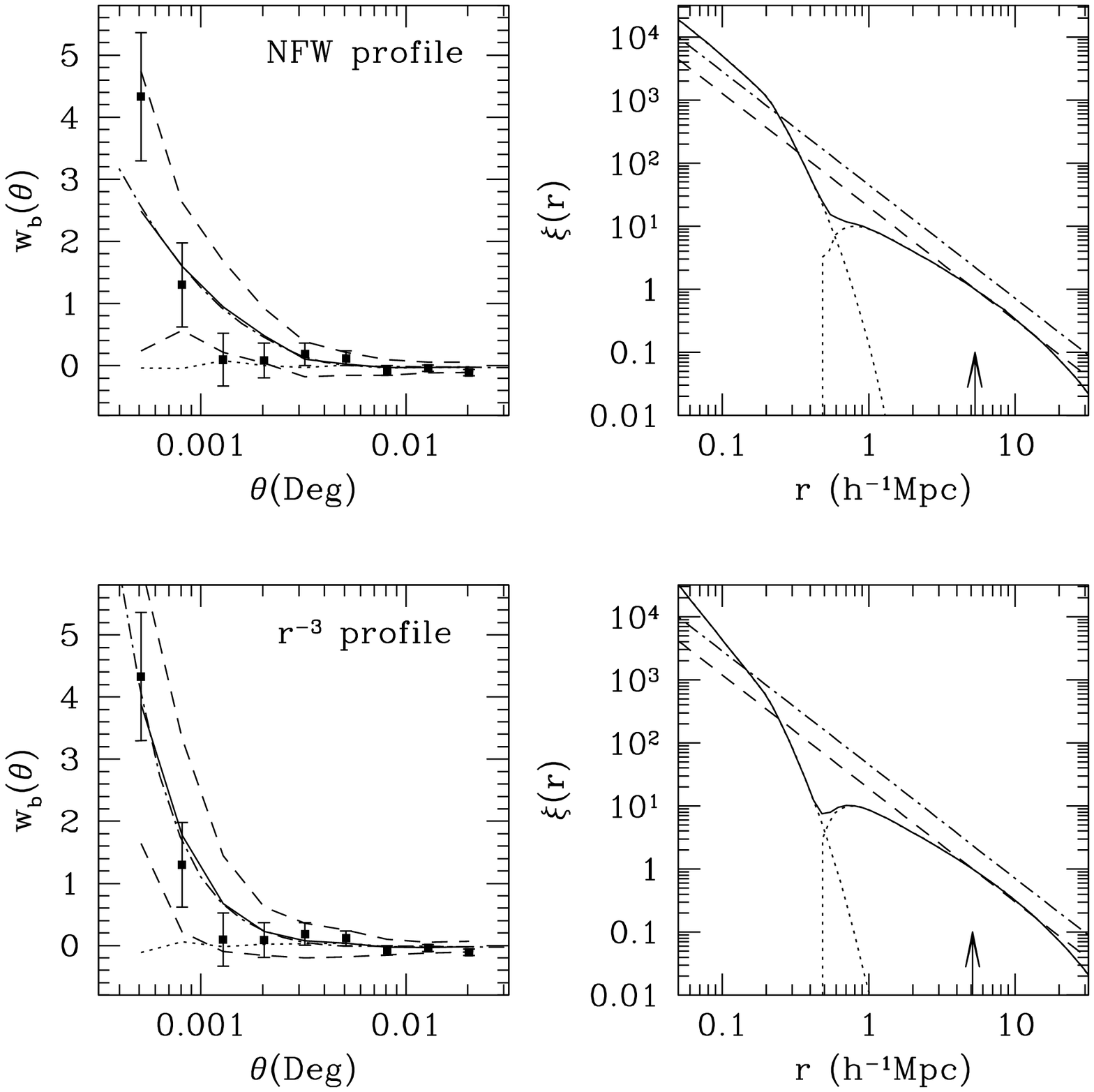}
\caption[]{Fitting results, sample variance from mock catalogs, and 
comparison for different assumptions about the galaxy distribution 
profile within halos. The nearest-integer distribution is used. 
{\it Top :} Galaxies distributed 
according to the NFW profile and $(M_{\rm min},M_1,\alpha)=(6.3\times 
10^{11} h^{-1}M_{\odot}, 2.9\times 10^{11} h^{-1}M_{\odot}, 0.45)$.
{\it Bottom :} Galaxies follow an $r^{-3}$ distribution profile and
$(M_{\rm min},M_1,\alpha)=(5.8\times 10^{11} h^{-1}M_{\odot}, 2.7\times 
10^{11} h^{-1}M_{\odot}, 0.38)$.  Left panels show the measured angular 
correlations $w_b(\theta)$. The solid, dashed, and dotted lines are the 
mean, 1$\sigma$ scatter about the mean, and the two-halo pair contribution, 
respectively, measured from mock galaxy catalogs generated through populating 
$z=2.97$ halos from the GIF simulation (see the text). The dot-dashed line 
is the analytic prediction of $w_b(\theta)$. Data points with error bars 
are from \citet{Daddi03}. The right panels show the corresponding real-space 
two-point correlation functions. The dotted lines are the one-halo and 
two-halo terms. Arrows indicate $r_0$ where $\xi(r_0)=1$. Two power law 
curves, $(r/r_0)^{-1.8}$ ({\it dashed curve}) and $(r/8.3\hMpc)^{-1.8}$ 
({\it dot-dashed curve}), 
are also plotted for comparison.
 }
\end{figure}
Perhaps the most important lesson is that the observed angular 
correlation signals are dominated by the one-halo term, where the
two galaxies of each pair are from one single halo. This can be
seen in the top left panel of Figure~2, where the dotted line shows 
the two-halo term. The contribution from the two halo term becomes comparable 
to that from the one-halo term only on angular scales greater than 
$\sim 0^\circ_\cdot005$ (also see the bottom left panel of Figure~1). 
As mentioned 
in \citet{Daddi03}, the estimated angular correlation at the smallest 
angular scale is mainly from a few triplets. The redshift distribution
of the galaxy sample provides further evidence. There are 
many spiky structures in the redshift distribution (\citealt{Daddi03}).
Since the largest projected separation in the field of view is about 
$3\hMpc$,  galaxies within the same spike are most likely to be 
physically close, and thus they have a high probability of being located 
in the same halo. 

Domination of the signal by the one-halo term has several implications. 
The HOD model generically predicts that the correlation function is not 
strictly a power law (see \citealt{Berlind02}). Instead, there should be 
a transition region from the one-halo--dominated regime at small scales to 
the two-halo--dominated regime at large scales. For the real-space 
correlation function, such a transition happens around the virial radius 
of the largest halos, $2-3\hMpc$ at $z=0$. Recent results from SDSS 
have revealed a statistically significant departure from a power law in the 
two-point correlation function, which can be well explained within the 
framework of the HOD (\citealt{Zehavi04}). Two-point correlation functions 
measured from other surveys, such as 2dFGRS and APM, also show such a 
departure (e.g. \citealt{Hawkins03,Padilla03}). 

The two-point correlation function predicted by the model that fits the 
angular correlation function in this paper shows a prominent departure 
from a power law (Figures~1 and 2), which is also reflected in the 
predicted angular correlation function (Figure~1). [Note that the excellent 
agreement between the numerical and analytic calculations of $w_b(\theta)$ 
demonstrates that any artifacts of our approximate treatment of halo 
exclusion are negligible in comparison to the observational error bars.]
Since the one-halo term is related to the distribution of galaxies within 
halos and the two-halo term is mostly determined by the halo-halo distribution, 
the amplitude and slope of the two terms may differ from each other 
substantially. One should therefore be cautious about inferring the 
correlation length by assuming a pure power law in the correlation function. 
In \citet{Daddi03}, a power law with an index $-1.8$ for the real-space 
correlation function, corresponding to an index of $-0.8$ for the angular 
correlation function, is assumed, and a high correlation amplitude 
(correlation length $\sim 8\hMpc$) is found. This strong correlation 
is unlikely to be related to the real correlation between halos, since, 
as we show here, the statistically significant signal is dominated by 
galaxy pairs within halos. The distribution of galaxies within halos does 
not tell how galaxies cluster on large scales, and the correlation length 
is overestimated because of the high amplitude of the one-halo term. In fact, 
from the fitting model, the correlation length where $\xi(r)=1$ can be as 
low as $\sim 5\hMpc$ (Figure~2), which is in a good agreement with the
result of \citet{Kravtsov04} for a subhalo sample of comparable number 
density. The mystery about the strong clustering 
in this particular sample then disappears. Our explanation of this mystery
is, in some sense, a quantitative version of the speculation of \citet{Daddi03}
that the strong clustering signal is an effect of ``excess" small-scale 
clustering.
 
We note that although the fit to the data can be regarded as 
acceptable, the third and the fourth data points are well below the
prediction. This may be of no significance considering the large sample
variance. Nevertheless, it is interesting to ask what the cause may be if 
this discrepancy is real. The low amplitude of these two data points may
be a hint that the one-halo term drops faster than in our model, which 
means that the distribution of galaxies within halos is more concentrated 
than the NFW profile we use. As an alternative model, we first doubled the 
concentration parameter of the galaxy distribution profile within each 
halo, thus making the galaxies more centrally concentrated than the dark 
matter, but this change is not adequate to match the observed $w_b(\theta)$
in the third and fourth angular bins. It thus implies that the distribution 
profile of galaxies is steeper than the NFW profile. As a more extreme 
alternative, we considered an $r^{-3}$ profile, with a flat core at 
$r<0.1\Rvir$ to make the pair distribution finite. The bottom panels 
of Figure~2 show the resultant model fitting and the sample variance 
estimated from mock catalogs generated using halos in the GIF simulation. 
The steeper galaxy distribution yields a better fit to the third and 
fourth data points.  As before, HOD parameters remain poorly constrained. 
With the current sample size, the preference for $r^{-3}$ profiles over
NFW profiles is not highly significant, but the low amplitude of the
third and fourth data points could be a hint that the red galaxies in 
the sample of \citet{Daddi03} are centrally concentrated within their parent
halos, analogous to the morphological segregation observed in present-day
clusters (e.g., \citealt{Oemler74,Melnick77,Dressler80,Adami98}).

Although HOD parameters are loosely constrained, the cutoff mass 
$M_{\rm min}$ in all the models shown in the figures roughly remains 
constant, $\sim 6\times 10^{11} h^{-1}M_\odot$. The approximate 
constancy of $M_{\rm min}$ mainly comes from the constraint of the 
galaxy number density and the steep drop of the halo number density 
toward higher halo masses. For example, the cumulative number density 
of halos drops from $\sim 5\times 10^{-3} h^3{\rm Mpc}^{-3}$ to 
$\sim 3\times 10^{-3} h^3{\rm Mpc}^{-3}$ as the minimum halo mass changes
from $5\times 10^{11} h^{-1}M_\odot$ to $7\times 10^{11} h^{-1}M_\odot$.
With the galaxy number density fixed, a large change in $M_{\rm min}$ is 
hard to compensate with changes in other HOD parameters. Although the
sharp cutoff in $\Navg(M)$ that we have assumed in this paper is an 
idealization, the derived value of $M_{\rm min}$ should still give an 
approximate indication of the characteristic minimum masses of halos 
that host the red galaxies.  In our successful models, the mean occupation 
number at $M_{\rm min}$ is above 1 (i.e., $M_{\rm min}>M_1$). This 
suggests that the red galaxies arise preferentially in groups and clusters 
(see \citealt{Moustakas02}), which may be a signature of an environmental 
effect on color. However, since $M_{\rm min} < M_2$, where $M_2$ is the 
mass of the halo that on average contains two red galaxies, there are 
still single-occupancy halos as the nearest-integer distribution is taken
into account. For example, in the model with $\alpha=0.45$
and $M_{\rm min}=6.3\times 10^{11} h^{-1}M_\odot$, 
$M_2 = 1.4\times 10^{12} h^{-1}M_\odot$, and about 11\% of the galaxies
are the sole occupants of their halos.

There are several hints that the red galaxies of the \citet{Daddi03} sample
have clustering properties different from those of the UV-selected galaxies
(e.g., LBGs) at the same redshift. The first hint comes from 
the $M_{\rm min} > M_1$ result mentioned above. Using similar kinds of 
HOD models (although assuming a pure power law with a low mass cutoff, 
with no single occupancy ``plateau"), \citet{Bullock02} and 
\citet{Moustakas02} find $\Mmin \sim 10^{10} h^{-1}M_\odot \ll M_1$ 
for LBGs at $z\sim 3$, implying that most LBGs are the sole occupants of 
their parent halos. By contrast, our model fits imply that most red 
galaxies reside in groups of two or more, similar to the results of 
\citet{Moustakas02} for local giant elliptical galaxies and $z\sim 1$ EROs 
(for which
they find a trend of $\Mmin > M_1$). A second hint is from the correlation 
length itself, which is still $\sim 5\hMpc$ in our models.  The correlation
length of UV-selected LBGs in the spectroscopic sample of \cite{Adelberger03}
is only about $4\hMpc$, and it appears to decrease for samples of lower
luminosity threshold and higher space density \citep{Giavalisco01}.
A final hint comes from the behavior of clustering on small scales.
\cite{Porciani02} find that the angular correlation function of $z\sim 3$
LBGs drops at separations of less than $30''$, while the angular clustering
of the red galaxies seems, if anything, to be exceptionally strong at the
smallest angular scales. To illustrate these differences between the red 
galaxies and the UV-selected LBGs, we show in Figure~1 what the HOD 
and correlation functions would look like for an LBG sample that has the same 
number density as the red galaxies. For this purpose, we start from the 
HOD model result of $z\sim 3$ LBGs by \citet{Moustakas02}, which is consistent 
with that of \citet{Bullock02}, and change their HOD parameters a little 
bit to match the number density here. The mean occupation function has
a power-law form $\Navg(M)=(M/M_1)^\alpha$ with a low-mass cutoff $\Mmin$.
We adopt $\Mmin=1.4\times 10^{10} h^{-1}M_\odot$, 
$M_1=4.0\times 10^{12} h^{-1}M_\odot$, $\alpha=0.8$, and a nearest-integer 
distribution. The dashed line in the top right panel of Figure~1 shows this
mean occupation function, where we can clearly see that unlike the red 
galaxies, relatively fewer LBGs reside in groups. This leads to 
a lower small-scale clustering amplitude and a lower correlation length with
respect to the red galaxies (see the dashed curves in the left panels of 
Figure~1). For a more consistent comparison between the red galaxies and the 
LBGs, we need detailed HOD modeling of the LBGs, which is out of the scope of 
this paper. The exercise here is to simply illustrate the differences in the 
clustering properties of the red galaxies and the LBGs, as noticed by 
\citet{Roche02}, \citet{Roche03}, and \citet{Daddi03} from angular clustering
measurements.
 
\section{Conclusion and Discussion}

In this paper, we have presented an HOD model of the observed strong clustering 
of a population of red galaxies at $z\sim 3$ analyzed by \citet{Daddi03}. 
Fitting the angular correlation data by assuming the real-space correlation 
to be a power law with the form $(r/r_0)^{-1.8}$, \citet{Daddi03} find the 
correlation length $r_0\sim 8\hMpc$, which would imply that galaxies reside 
in rare halos with $M\ge 10^{13}h^{-1}M_\odot$ and which would require a 
very large occupation number in each halo to account for the observed 
galaxy number density. Our HOD modeling shows that the angular clustering 
data can be explained by a less surprising model, e.g., with a cutoff at 
$6.3 \times 10^{11} h^{-1}M_\odot$ and mean galaxy occupation number 
$\Navg(M)=1.4(M/6.3 \times 10^{11} h^{-1}M_\odot)^{0.45}$ above this 
cutoff. Artificial galaxy catalogs constructed with this HOD show that
sample variance increases error bars by $\sim 50$\% over those estimated 
by \citet{Daddi03}, which (as they noted) did not take sample variance 
into account. 

There is degeneracy between HOD parameters $M_1$ and $\alpha$. However,
the characteristic minimum mass of halos that can host the red galaxies
seems to be around $6 \times 10^{11} h^{-1}M_\odot$. Results from our
modeling suggest that the red galaxies are a different population from LBGs.

HOD parameters are not tightly constrained by the data, but in all cases
the significantly non-zero points from \citet{Daddi03} are in a regime 
dominated by pairs within single halos. The amplitude of the correlation 
function in the two-halo regime is below an $r^{-1.8}$ power law extrapolation 
of that of the one-halo regime, which is why lower mass halos are acceptable. 
Thus, if our explanation is correct, surveys with larger area should 
show weaker correlations than this $r^{-1.8}$ extrapolation. The 
correlation length predicted by our model can be as low as $\sim 5 \hMpc$,
a prediction that can be tested by larger area surveys. 

Obtaining good constraints on HOD parameters will require samples large
enough to accurately probe the two-halo regime as well as the one-halo regime. 
Spectroscopic surveys are also of importance since they allow one to 
measure galaxy clustering in redshift space (in addition to more accurate 
measurements of the projected clustering). With a good understanding of the
velocity field of halos, the clustering in redshift space would at least
provide a consistency check for the HOD model. With wider angle space-based 
surveys such as GOODS\footnote[1]{See http://www.stsci.edu/science/goods.}
and ambitious infrared follow-up programs like the FIRES project 
\citep{Franx00,Daddi03}, the necessary data should become available in 
the next several years. This will provide valuable constraints on the host 
halos of red high-$z$ galaxies and clues to their formation histories.

\acknowledgments

We thank Emanuele Daddi not only for providing us the observation geometry 
and the angular clustering data used in this paper but also for his useful 
comments and suggestions. We also thank David Weinberg for helpful 
discussions, for his encouragement on this work, and for his valuable 
comments that improved the paper. This work was supported by NSF grant 
AST 00-98584 and by a Presidential Fellowship from the Graduate School of 
the Ohio State University.

\end{document}